\documentclass[prb,aps,showpacs,twocolumn,preprintnumbers,amsmath,amssymb,superscriptaddress,nofootinbib]{revtex4-1}

\usepackage{amsmath,amssymb,amsfonts}
\usepackage{graphicx}
\usepackage[colorlinks=True,linkcolor=black,citecolor=blue,urlcolor=blue]{hyperref}
\usepackage{xcolor}
\usepackage{chngcntr}
\usepackage{braket}
\usepackage{hyperref}
\usepackage{nicefrac}
\usepackage{bm}
\usepackage{float}
\usepackage{ulem}
\usepackage[export]{adjustbox}

\usepackage{booktabs}

\newcommand{\bs}[1]{\boldsymbol{#1}}

\begin{document}

\title{Magnetic injection photocurrents in valley polarized states of twisted bilayer graphene}

\author{Fernando Peñaranda*}
\affiliation{Donostia International Physics Center, P. Manuel de Lardizabal 4, 20018 Donostia-San Sebastian, Spain}
\author{Héctor Ochoa}
\affiliation{Department of Physics, Columbia University, New York, NY 10027, USA}
\author{Fernando de Juan}
\affiliation{Donostia International Physics Center, P. Manuel de Lardizabal 4, 20018 Donostia-San Sebastian, Spain}
\affiliation{IKERBASQUE, Basque Foundation for Science, Plaza Euskadi 5, 48009 Bilbao, Spain}

\date{\today}
%%%%%%%%%%%%%%%%%%%%%%%%%%%%%%%%%%%%%%%%%%%%%%%%%%%%%%%%%%%%%%%%%%%%%%%%%%%%%
\begin{abstract}
Magic-angle twisted bilayer graphene displays a complex phase diagram as a function of flat band filling, featuring compressibility cascade transitions and a variety of competing ground states with broken spin, valley, and point group symmetries. Recent THz photocurrent spectroscopy experiments have shown a dependence on the filling which is not consistent with the simplest cascade picture of sequential filling of equivalent flat bands. In this work, we show that when time-reversal symmetry is broken due to valley polarization, a magnetic injection photocurrent develops, which can be used to distinguish different spin-valley polarization scenarios. Using the topological heavy fermion model, we compute both shift and injection currents as a function of filling. We argue that current experiments can be used to determine the spontaneous valley polarization.  
\end{abstract}
\maketitle

\section{Introduction} Magic-angle twisted bilayer graphene (TBG)~\cite{Bistritzer11} is a unique correlated electron system displaying a wide variety of broken-symmetry ground states~\cite{Cao2018Insulators,Lu19,Kerelsky19,Xie19,Stepanov21,Sharpe19,Serlin20,Cao21Nematicity,Tseng22} and unconventional superconductivity~\cite{Cao18SC,Yankowitz19,Oh21}. Its low-temperature phase diagram as function of flat-band filling $\nu$ remains a debated subject, where the %enhanced
Coulomb interaction favors competing states that originate from the partial filling of spin ($\uparrow/\downarrow$) and valley ($\pm$) degenerate flat bands and with strong dependency on substrate effects and strain. While valley-coherent states~\cite{Bultinck20,Kwan21} are competitive ground states in the absence of a substrate, the alignment with the hexagonal boron nitride (hBN) strongly favors polarized states~\cite{Zhang20BN,Kwan21,Liu21}. These are characterized by what flavors %: spin ($\uparrow/\downarrow$) and valley ($\pm$), 
are filled, each having a Chern number given by the product of the sublattice and valley eigenvalue\cite{Xie20,Zhang19,Bultinck20AHE,Liu21}. Odd fillings always leads to a spin- and valley-polarized Chern insulator with $|C|=1$, while for filling $\nu=2$, a valley-polarized quantum anomalous Hall (QAH) state with $C=2$, $\left|\uparrow+,\downarrow + \right>$, is nearly degenerate with a ferromagnetic valley Hall (VH) $\left|\uparrow+,\uparrow -\right>$ and spin-valley Hall (SVH) $\left|\uparrow+,\downarrow -\right>$ states with zero Chern number~\cite{Kang19,Liu21,Xie21}. Other broken-symmetry states like nematic semimetals \cite{Kang20,Liu21Nematic} and stripe phases \cite{Kang20,Xie23} are also predicted, as well as symmetry-preserving Mott and symmetric Kondo~\cite{Chou23,Hu23,Zhou24,Rai24} states, which may be realized at higher temperatures.

Of particular interest is the potential breaking of time-reversal symmetry in some of the these low-temperature ground state candidates, an experimental feature observed at odd fillings $\nu=1,3$ in the electron~\cite{Sharpe19,Serlin20,Stepanov21} and hole sides~\cite{zhang24HoleSide}, and under some circumstances near $\nu=2$~\cite{Tseng22}, displaying %quantum anomalous Hall (
QAH effect signatures. Experimentally, these states appear to emerge from a parent high-temperature state where the different flavors are filled sequentially as a function of carrier density, causing a periodic reset of the chemical potential known as cascade behavior, observed both in local tunneling~\cite{Wong20,Nuckolls20,Choi21,Choi21NP} and compressibility~\cite{Zondiner20,Rozen21}. Since the cascade behavior remains up to much higher temperatures (20-30 K)~\cite{Zondiner20,Rozen21,Choi21NP} than those where insulating states develop at integer fillings, the cascade itself does not necessarily signal symmetry breaking~\cite{Datta23,Calderon24}, while the QAH effect does.

Photocurrent THz spectroscopy~\cite{Otteneder20,Hubmann22,Ma22} is 
a unique optical probe~\cite{Hesp21,Sunku21} to examine the symmetry and quantum geometry of TBG flat bands~\cite{Liu20Anomalous,Kaplan22,Zhang22Correlated,Chaudhary22,Arora21,Penaranda24,Postlewaite24}, and can offer a unique insight regarding the breaking of time reversal symmetry complementary to standard transport probes. %, \blue{beyond QAH measurements}. 
 Recent experiments have shown photocurrents are sensitive to the cascade behaviour~\cite{Kumar25NM}, albeit with low temperature sign-changing features which are inconsistent with a naive periodic resetting of the chemical potential. This suggests that photocurrents are sensitive to the flavor polarization realized at each filling. 

\emph{In this work}, we propose that a particular mechanism known as magnetic injection current~\cite{Zhang19,Fei20,Wang20,Ahn20,Holder20,Watanabe21,Watanabe21b,Merte21,Okumura21}, which switches sign with magnetization, is the dominant contribution to the photocurrents in valley polarized states, and may be responsible for the observations. Using the topological heavy fermion model (THFM)~\cite{Song22} -see sketch in Fig.~\ref{fig:sketch}a)-, we compute both shift and injection currents for continuously varying flat band filling in the Hartree-Fock approximation, and discuss how these two contributions can be separated from symmetry considerations. We demonstrate that the magnetic injection current can be used to discriminate between low-temperature polarized ground states, distinguishing topological QAH from trivial VH or SVH phases.

This paper is organized as follows: Sec.~\ref{Photovoltaiceffects} presents the symmetry analysis of photogalvanic responses, with Sec.~\ref{photononinteracting} devoted to the specifics of the photogalvanic response in TBG. The theoretical model is presented in Sec.~\ref{model}, with Sec.~\ref{THFM} discussing the THFM, Sec.~\ref{substrateeffects} devoted to the substrate effects with an explicit construction of the $z$-position operator matrix elements projected onto the THFM degrees of freedom, and Sec.~\ref{correlations} devoted to the Hartree-Fock solution. The photocurrent spectra in the valley polarized states are reported in Sec.~\ref{photospect}, and the discussion follows in Sec.~\ref{conclusion}.

\section{Photogalvanic effects in TBG}\label{Photovoltaiceffects}

\subsection{Shift and injection currents}

In non-centrosymmetric systems in the presence of constant light irradiation, a DC photocurrent is generated with the general form 
\begin{align}
J_i &= \sigma_{ijk}(E_jE_k^* + E_j^*E_k) + i\eta_{ijk}(E_jE_k^* - E_j^*E_k),
\end{align}
where $\sigma_{ijk} =\sigma_{ikj} $ denotes the linear photogalvanic effect (LPGE), while $\eta_{ijk} = -\eta_{ikj}$ denotes the circular photogalvanic effect (CPGE). Both effects are generated by two main mechanisms known as shift and injection currents~\cite{Aversa95}. Focusing only on LPGE, relevant to current experiments, we can write $\sigma_{ijk}= \sigma_{ijk}^{\rm sh} +\sigma_{ijk}^{\rm inj}$, where 
\begin{align}
\sigma_{ijk}^{\rm sh} &= \frac{\pi e^3}{2\hbar^2} \int_{\boldsymbol{k}} \sum_{nm} f_{nm}{\rm 
Im}[r^k_{nm;i}r^j_{mn}-r^k_{nm}r^j_{mn;i}] \delta(\omega - \omega_{nm}),\label{shift}\\
\sigma_{ijk}^{\rm inj} &= \tau \frac{\pi e^3}{\hbar^2}\int_{\boldsymbol{k}} \sum_{nm} f_{nm} \partial_{k_i}\omega_{nm} {\rm Re}[r_{nm}^jr_{mn}^k] \delta(\omega - \omega_{nm}). \label{maginj}  
\end{align}
Here $\tau$ denotes the scattering time, $\int_{\boldsymbol{k}} =\int \tfrac{d^2k}{(2\pi)^2}$, $f_{nm}$ is the difference between the Fermi-Dirac occupation of states $|n\rangle$ and $|m\rangle$ ($\boldsymbol{k}$ dependence is implicit), $\omega_{nm} = (E_n-E_m)/\hbar$ with $E_n$ the energy of state $n$, $r_{nm}^i = \left<n\right| i\partial_{k_i} \left|m\right>$ for $i=x,y$, and  $r^i_{nm;j}$ is the generalized derivative computed with the aid of the standard sum rule~\cite{Aversa95}.
\begin{figure}[t]
    \centering
    \includegraphics[width=0.9\linewidth]{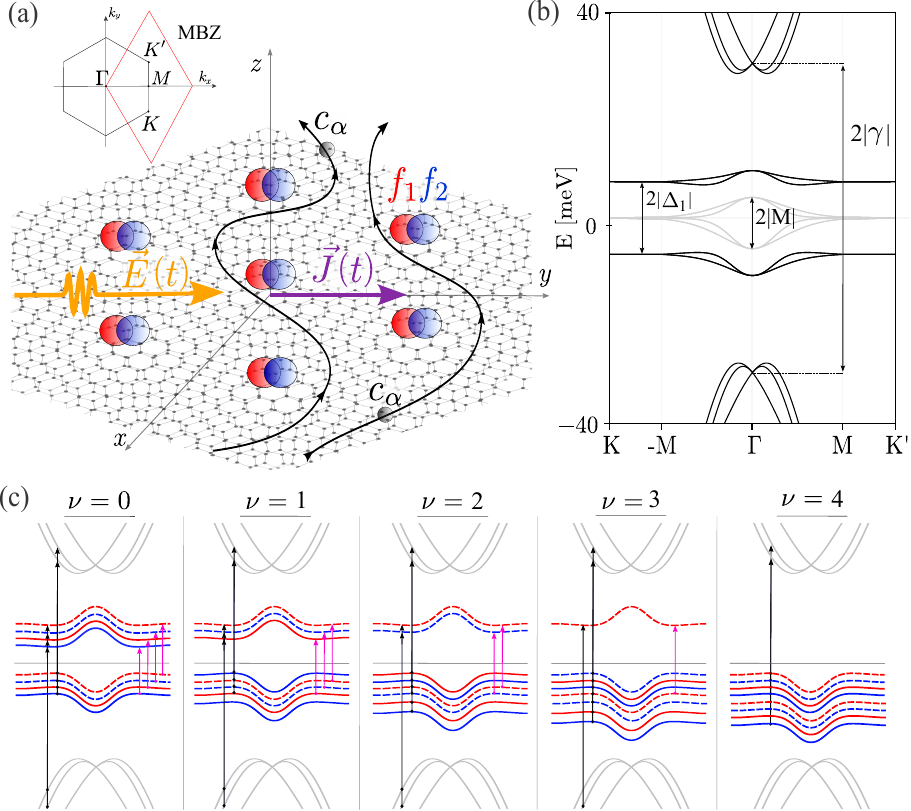}

    \caption{(a) Sketch of a TBG flake in the presence of an incident in-plane electric field $\boldsymbol{E}(t)$ and the resulting induced non-linear current $\boldsymbol{J}(t)$. The two local $f$ orbitals (per valley and spin) in the heavy fermion description of TBG are depicted in red and blue at the AA positions of the moiré pattern, whereas the $c$ electrons of the delocalized conduction bands are shown in black. The high symmetry points of the moiré Brillouin zone (MBZ) are shown in the figure inset. (b) Bands of the non-interacting THFM close to magic angle in the absence (gray) and presence (black) of a layer-even sublattice mass. (c) Schematic of the allowed optical transitions between occupied and unoccupied bands at integer $\nu$. Flat to flat (FF) transitions are depicted in pink and flat to dispersive (FD) in black. Solid and dashed, and red and blue encode different valley and spin flavors, respectively.}
    \label{fig:sketch}
\end{figure}
The shift contribution $\sigma_{ijk}^{\rm sh}$ is a $\mathcal{P}$-odd, $\mathcal{T}$-even response (where $\mathcal{P}$ is inversion and $\mathcal{T}$ is time-reversal symmetry), while the injection contribution $\sigma_{ijk}^{\rm inj}$ is a $\mathcal{T}$-odd, $\mathcal{PT}$-even response which occurs only in magnetic systems and switches sign when magnetization is reversed. 

In TBG, where single-electron excitations can be described in terms of independent valleys related by time-reversal symmetry, the linear injection current can be seen as a valley photocurrent~\cite{Golub11,Kaplan22}, which becomes a true charge photocurrent once time-reversal symmetry is broken and the contributions from the two valleys do not cancel each other. Since only the injection current depends on $\tau$, in clean magnetic systems with large $\tau$, the injection current dominates over the shift current.

\subsection{Symmetry analysis of the photocurrents in TBG}\label{photononinteracting}

TBG has lattice point group $D_6$ generated by twofold and threfold rotations along the vertical axis, $C_{2z}$, $C_{3z}$, and a twofold rotation along an axis within the plane, $C_{2x}$. Experimentally, photocurrents in TBG were reported at normal incidence \cite{Kumar25NM}, i.e. $\sigma_{ijk}$ with $i,j,k = x,y$. These components are forbidden by $C_{2z}$ symmetry,~\cite{Penaranda24}  indicating that it has been broken. In addition, photocurrents are observed up to 60 K \cite{Kumar25NM}, which is much higher than the temperature scale of correlation effects. This implies that the breaking of $C_{2z}$ is not spontaneous but external and originates from the alignment with  hBN.

The hBN lattice structure breaks $C_{2z}$ but preserves $C_{3z}$. When TBG is placed on hBN, the average effect of the substrate is therefore to break $C_{2z}$ as well as the in-plane two-fold axes $C_{2x}$ and $C_{2y}=C_{2x}\times C_{2z}$ which interchange the two graphene layers. These effects are reflected in the electronic degrees of freedom as three different perturbations with different transformation properties under the twofold axes $C_{2x}$, $C_{2y}$, $C_{2z}$, which in turn affect the different photocurrent components differently~\cite{Penaranda24}. The dominant effect is an electric potential which is opposite in the two carbon sublattices within a layer (a sublattice potential) and is the same on both layers, which we denote as $\Delta_1$. This potential transforms as the irrep $B_2$ (even under $C_{2y}$). The next effect is a sublattice potential $\Delta_2$ which is opposite in the two layers and transforms as $B_1$ (even under $C_{2x}$). Finally, there is a small interlayer potential $\Delta_3$ with symmetry $A_2$ (even under $C_{2z}$) which is often neglected.

When $C_{2z}$ is broken by these sublattice potentials but $C_{3z}$ is preserved, two components of the LPGE become allowed~\cite{Liu20Anomalous,Kaplan22,Zhang22Correlated,Chaudhary22}, which also transform as irreps of the point group: $\sigma_{yyy}=-\sigma_{yxx}=-\sigma_{xxy}$, which transforms as $B_1$ and $\sigma_{xxx}=-\sigma_{xyy}=-\sigma_{yxy}$ which transforms as $B_2$. Since $\mathcal{T}$ is preserved and only shift contributions are present, to leading order in the substrate perturbations, $\sigma_{yyy}^{\rm sh}$ is proportional to the layer even part of the sublattice potential $\Delta_1$, while $\sigma_{xxx}^{\rm sh}$ is proportional to the layer-odd part $\Delta_2$ \cite{Penaranda24}.

Spontaneous breaking of time-reversal and valley symmetries in the correlated states lifts additional constraints in the allowed photocurrent components. The symmetry group of a single valley is generated by $C_{3z}$, $C_{2x}$ and $C_{2z}\mathcal{T}$, while $C_{2z}$, $C_{2y}$ and $\mathcal{T}$ interchange valleys. Therefore, valley polarization can be seen as a time-odd $B_1$ perturbation, which enables the time-odd $B_1$ photocurrent component $\sigma_{yyy}^{\rm inj}$. In the general case where both the substrate and valley polarization are present, both $\sigma_{xxx}$ and $\sigma_{yyy}$ are finite, but to leading order they remain approximately proportional to the corresponding perturbations with the same symmetry.

\section{Electronic spectrum of TBG}\label{model}

To compute the photocurrent spectrum in valley-polarized states we consider an interacting electron Hamiltonian consisting of four terms,\begin{align}
\hat{H}=\hat{H}_0+\hat{H}_{U}+\hat{H}_{J}+\hat{H}_{\textrm{subs}}.
\end{align}
Next, we introduce each of these terms and the Hartree-Fock solutions as a function of electron filling.

\subsection{The Topological Heavy Fermion Model}\label{THFM}
We start with the Hamiltonian of the THFM~\cite{Song22}, which is a faithful representation of the continuum model \cite{dosSantos12,Bistritzer11} that separates the degrees of freedom into localized, strongly interacting states $f$ and extended, topological states $c$ %with a metallic quadratic band touching 
(see Fig. \ref{fig:sketch}(a)). The single-particle THFM Hamiltonian is given by:
\begin{align}
    \hat H_0 & = \sum_{|\boldsymbol k|<\Lambda_c} \sum_{a, a', \eta, s} \left( H_{aa'}^{(cc,\eta)} (\boldsymbol k) - \mu \delta_{aa'}  \right) \hat c^\dagger_{\boldsymbol k a \eta s} \hat c_{\boldsymbol k a' \eta s} \nonumber \\ &
    - \mu \sum_{\alpha\eta s} \sum_{\bs R} \hat f^\dagger_{\bs R \alpha \eta s} \hat f_{\boldsymbol R \alpha \eta s}  + \frac{1}{\sqrt{N}} \times \\
    & %\times \\&\times
    \sum_{\substack{|\boldsymbol k|<\Lambda_C, \bs R \\\alpha a \eta s}} \left[ e^{i\boldsymbol k \cdot \boldsymbol R - \frac{|\bs{k}|^2 \lambda^2}{2}} H_{\alpha a}^{(fc, \eta)}(\bs k) \hat f^\dagger_{\bs R \alpha \eta s} \hat c_{\boldsymbol{k} a \eta s} + H.c.\right], \nonumber \label{H0}
\end{align}%\end{widetext}
\interfootnotelinepenalty=10000
where $\hat c^\dagger_{\boldsymbol{k} a s \eta}$ ($\hat c_{\boldsymbol{k} a s \eta}$) creates (destroys) a $c$ electron in conduction band $a \in \{1,2,3,4\}$ with spin $s$, valley $\eta$, and momentum $\boldsymbol k$ smaller than the $c$-electron momentum cutoff $\Lambda_C$\footnote{To ensure periodicity in the moiré Brillouin zone in bandstructure calculations, one can safely assume the $\Lambda_C \rightarrow \infty$ without affecting the low energy physics and, then, truncate the plane-wave basis expansion in moiré reciprocal lattice vectors to the first shell.}, while $\hat f^\dagger_{\boldsymbol{R} \alpha s \eta}$ ($\hat f_{\boldsymbol{R} \alpha s \eta}$) creates (destroys) an $f$ electron with orbital $\alpha \in \{1,2\}$ at the moiré unit cell located at position $\boldsymbol R$, with spin $s$ and valley $\eta$. $N$ denotes the number of moiré unit cells, $\mu$ the chemical potential, and $\lambda = 0.3375a_M$ is a dampening factor proportional to the spread of the local orbitals, being $a_M$ the moiré lattice constant. $H^{(cc, \eta)}$ is the non-interacting Hamiltonian for the $c$ electrons:
 \begin{align}
    H^{(cc, \eta)} = \begin{pmatrix}
        0_{2\times2} & v_\star (\eta k_x \sigma_0 + ik_y\sigma_z)\\
        v_\star (\eta k_x \sigma_0 - ik_y\sigma_z) & M \sigma_x
    \end{pmatrix},
\end{align}
and
\begin{align}
    H^{(fc, \eta)}(\bs k) = [\gamma \sigma_0 + v_\star' (\eta k_x \sigma_x + k_y \sigma_y), 0_{2\times 2}],
\end{align}
with $\sigma_{0,x,y,z}$ the Pauli matrices in orbital space, captures the  $f$-$c$ hybridization responsible for two isolated flat-bands with bandwidth $2|M|$ separated from the dispersive bands by $|\gamma| - |M|$. The resulting bandstructure is shown in Fig.~\ref{fig:sketch}(b). %which corresponds to an exact mapping to the low-energy sector of the continuum model. 
$v_\star$, $v_\star'$, $M$, and $\gamma$ values are obtained from the continuum model parameters at a twist angle $\theta = 1.05$~\cite{Song22}.

Adding interactions to the THFM is achieved by projecting the screened Coulomb potential into the $f$ and $c$ degrees of freedom\cite{Song22}. This procedure leads to an interacting Hamiltonian in close resemblance to a generalized Anderson model with a density-density interaction term between $f$ electrons and parametrized by $U$ given by
\begin{align}
\hat{H}_U = \frac{U}{2} \sum_{\boldsymbol{R}}:\hat \rho^f_{\boldsymbol{R}}::\hat\rho^f_{\boldsymbol{R}}:, \label{HU_I}
\end{align}
and a ferromagnetic exchange coupling
\begin{align}
\hat{H}_J = -J \sum_{\boldsymbol{R}\boldsymbol{q}}\sum_{\mu \nu} \sum_{\chi = \pm} e^{-i\boldsymbol{q}\cdot \boldsymbol{R}} :\hat\Sigma_{\mu; \nu}^{f,\chi}(\boldsymbol R): :\hat \Sigma_{\mu ;\nu}^{c'',\chi}(\boldsymbol{q}):, \label{HJ_I}
\end{align}
with exchange constant $J$, that couples $f$ electrons and those $c$ electrons forming a $\Gamma_1\oplus\Gamma_2$ representation ($a = {3,4}$). %$\hat H = \hat H_0+\hat H_U+\hat H_J$ results in an interacting Hamiltonian which resembles a generalized Anderson model.
Other terms also resulting from the projection that leads to an effective shift in the bands were neglected\cite{Song22}. Colons above denote normal ordering, $\mu$ and $\nu \in \{0,x,y,z\}$, $\hat \rho^f_{\boldsymbol R} = \sum_{\alpha \eta s} \hat f_{\boldsymbol{R}\alpha \eta s}^\dagger \hat f_{\boldsymbol{R}\alpha \eta s} $, and $\hat \Sigma_{\mu \nu}^{(i,\chi)}$ with $i = \{f,c\}$ and $\chi = \pm$ as $8\times8$ matrices that become $U(4)$ moments in the flat-band limit realized at $M=0$. They read
\begin{align}
    \hat \Sigma_{\mu \nu}^{(f,\chi)} (\boldsymbol R) = \sum_{\substack{\alpha, \alpha'=1,2\\ \eta, \eta',s,s'}}\frac{\delta_{\chi, \eta (-1)^{\alpha-1}}}{2}A^{\mu; \nu}_{\alpha \eta s, \alpha' \eta' s'} \hat f^\dagger_{\boldsymbol{R} \alpha \eta s} \hat f_{\boldsymbol{R}\alpha' \eta' s'},
\end{align}
and
\begin{align}
    \hat \Sigma_{\mu \nu}^{(c,\chi)}(\boldsymbol q)  = \sum_{\substack{a, a'=3,4\\ \eta, \eta',s,s'}}\frac{\delta_{\chi, \eta (-1)^{a-1}}}{2N} B^{\mu; \nu}_{a\eta s, a' \eta' s'}\hat c^\dagger_{\boldsymbol{k}+\boldsymbol{q} a \eta s} \hat c_{\boldsymbol{k}a' \eta' s'},
\end{align}
with 
\begin{align}
A^{\mu;\nu} &= \{ \sigma_0\tau_0\varsigma_\nu,\sigma_y\tau_x\varsigma_\nu,\sigma_y\tau_y\varsigma_\nu,\sigma_0\tau_z\varsigma_\nu  \},\nonumber\\
B^{\mu;\nu} &= \{ \sigma_0\tau_0\varsigma_\nu,-\sigma_y\tau_x\varsigma_\nu,\sigma_y\tau_y\varsigma_\nu,\sigma_0\tau_z\varsigma_\nu  \}
\end{align}where $\tau_{0,x,y,z}$, $\varsigma_{0,x,y,z}$ act on valley and spin degrees of freedom. 

\begin{figure}[t]
    \centering
    \includegraphics[width=1\linewidth]{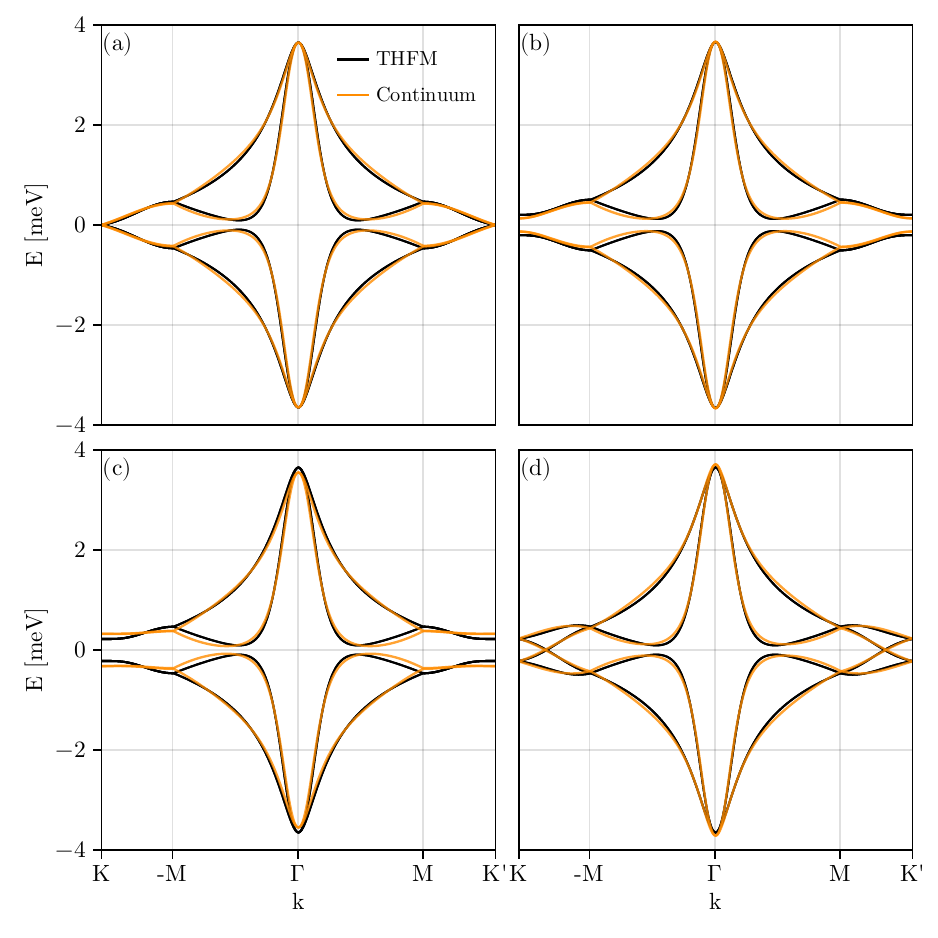}
    \caption{Bandstructure comparison between the non-interacting THFM and the continuum model in the absence (a) and presence of the three sublattice potentials in Eq~\eqref{subs}: $\Delta_1 = \Delta$ (b), $\Delta_2 = \Delta$ (c), and $\Delta_3 = \Delta$ (d), with $\Delta = 0.25$ meV. The continuum model parameters are: the twist angle $\theta = 1.05 \deg$, the intralayer hopping $t = -2.46575$ \text{eV}, and the interlayer hopping between AA and AB regions $t_{AA} = 0.078975 \text{eV}$, $t_{AB} = 0.0975 \text{eV}$, respectively. The THFM parameters $v_\star$, $v_\star'$, $M$, and $\gamma$ values are taken from Ref.~\cite{Song22}.} \label{figlatbands}
\end{figure}

\subsection{Substrate effects} \label{substrateeffects}

As explained in Sec. \ref{photononinteracting}, the leading effect of the substrate is to induce the three types of constant potentials $\Delta_i$ with symmetries $B_2$, $B_1$ and $A_2$, with moir\'e effects playing a secondary role \cite{Moon14,Long23}. Continuum model calculations thus often model the substrate by just including one or several of these potentials~\cite{Kaplan22,Chaudhary22,Kumar25NM}. For the THFM the form of these potentials has not been established yet, but as we show below it can be obtained by symmetry analysis or by direct projection from the continuum model. Either method reveals the substrate Hamiltonian to be
%Neglecting these moir\'e effects, the general effect of the substrate projected into the THFM degrees of freedom reads}
\begin{align}
\hat H_{\text{subs}} &= \sum_{\substack{\bs R\\ \alpha \alpha' s \eta}} \Big[ \Delta_1  \hat f^\dagger_{\bs R \alpha s \eta} \sigma^z_{\alpha \alpha'}\hat f_{\bs R \alpha' s \eta} \nonumber\\
&+ \sum_{i=1}^3    \hat f^\dagger_{\bs R \alpha s \eta} i\eta\xi(\Delta_2 \sigma^z_{\alpha \alpha'}+\Delta_3 \delta_{\alpha \alpha'}) \hat f_{\bs R + \bs a_i \alpha' s \eta}\Big],    \label{subs}
\end{align}
where $\boldsymbol{a}_i$ with $i\in \{1,2,3\}$ are the moir\'e lattice vectors and $\xi^{-1} = - \frac{1}{2\pi}\sin\frac{\pi}{\sqrt{3}}+\frac{1}{4\pi}\sin\frac{2\pi}{\sqrt{3}}$.

Note that the layer degree of freedom of the continuum model is lost in the projection to the THFM. Because of this, while the layer ever potential $\Delta_1$ admits the standard local representation as $f$-fermion orbital polarization $\sigma_z$ (transforming as $B_2$), the layer odd potentials $\Delta_2$ and $\Delta_3$ have become non-local.

The layer operator representation, which is equivalent to the $z$ component of the position operator $\hat r_z$, can be obtained in the symmetry approach as follows. $\hat r_z$ transforms as $A_2$ and is time even and particle-hole odd. It turns out that no local (scalar) irreducible representation within the space formed by the $f$ and $c$ electrons is compatible with these symmetries. However, it is possible to write a non-local term in the $f$ subspace that is consistent with the symmetries
\begin{align} \hat r^z = 2\xi \eta   \sum_{i \in{0,1,2}} \sin (\boldsymbol{k} \cdot \boldsymbol{a}_n) \hat f^\dagger_{\boldsymbol{k}, \alpha s \eta} \hat f_{\boldsymbol{k},\alpha s \eta}, \label{rz} \end{align}
where $\hat f^\dagger_{\boldsymbol{k},\alpha s \eta} = \frac{1}{\sqrt{N}}\sum_{\boldsymbol R} \exp(i\boldsymbol k \cdot \boldsymbol R) \hat f^\dagger_{\boldsymbol{R},\alpha s \eta}$ This operator, which can be interpreted as loop currents of the $f$ fermions, correctly captures the opposite energy shift at $K$ and $K'$ and changes signs at different valleys. Fourier transforming Eq.~\eqref{rz} we thus obtain the term for the $\Delta_3$ potential. To obtain the corresponding $\Delta_2$ we simply note that $B_1 = B_2 \times A_2$ so adding an extra $\sigma_z$ to this non-local operator produces the $\Delta_2$ potential.

The validity of the symmetry analysis for the derivation of $\hat r_z$, is further tested either by means of a numerical projection into the THFM $f-f$ subspace, and also by comparison between the continuum model and the THFM  bands, shown in Fig.~\ref{figlatbands} at a $1.05 \deg$ twist. In the absence of a substrate (a) and for each of the potentials in Eq.~\eqref{subs}: $\Delta_1 = \Delta$ (b), $\Delta_2 = \Delta$  (c), and $\Delta_3 = \Delta$ with $\Delta=0.25$ meV, the THFM bands (yellow) successfully captures the low-energy characteristics of the continuum model bands (black). The agreement found in panels (c) and (d) suggests that indeed Eq.~\ref{rz} is a faithful representation of $\hat r_z$.

Other non-local representations, acting on the $c-c$ or $c-f$ subspaces involving powers of $\mathbf{k}$, are also compatible with the symmetries of $r_z$, however, they are not expected to influence the flat bands, let alone shift their Dirac points. The reason is that at the $K/K'$ valleys they are $f$ polarized and their coupling with the $c$ electrons is strongly suppressed by a large energy scale. Therefore, these representations can be thought of as perturbative corrections to Eq.~\eqref{rz}, and thus negligible in the small $\Delta_2/|\gamma|$ limit.

\subsection{Hartree-Fock solutions} \label{correlations}
Hartree-Fock decoupling of the interacting terms in Eqs.~\eqref{HU_I}-\eqref{HJ_I} leads to:

\begin{align}
    \hat H_U & \rightarrow - \frac{N U}{2} \left( \nu_f^2 + 8\nu_f - \text{Tr}\left[ O^f O^f\right]\right) +U\sum_{\bs R} \sum_{\substack{\alpha s \eta \\ 
    \alpha' s' \eta'}} \nonumber\\
    &\left[ (\nu_f + \frac{1}{2}) \delta_{\alpha \alpha'}\delta_{s s'} \delta_{\eta \eta'} -  O^f_{\alpha s \eta, \alpha' s'\eta'} \right] \hat f^\dagger_{\bs R \alpha' s' \eta' }  \hat f_{\bs R \alpha s \eta} \nonumber \\
    &
    \label{HU}
\end{align}
and
\begin{align}
    \hat H_J &\rightarrow -\frac{J}{2} \sum_{|\bs k|<\Lambda_C} \sum_{\substack{\alpha s \eta \\ \alpha' s' \eta'}} \Bigg[(\eta \eta' + (-1)^{\alpha+\alpha'}) \times \nonumber \\
    & \times \left(O^f_{\alpha \eta s, \alpha' \eta' s'} -\frac{1}{2} \delta_{\alpha \alpha'} \delta_{s s'} \delta_{\eta \eta'}\right)\Bigg] \hat c_{\bs k, \alpha'+2, \eta' s'}^\dagger \hat c_{\bs k, \alpha+2, \eta s}
    \label{HJ}
\end{align}
where $O^f_{\alpha s \eta, \alpha' s' \eta'} = \langle \psi | \hat f^\dagger_{\bs R \alpha \eta s} \hat f_{\bs R \alpha' \eta' s'}|\psi\rangle$ denote the density matrix of the $f$ electrons in terms of the interacting mean-field ground state $\psi$, and $\nu_f = \text{Tr}[O^f] - 4$.
\begin{figure}[t]
    \centering
    \includegraphics[width=1\linewidth]{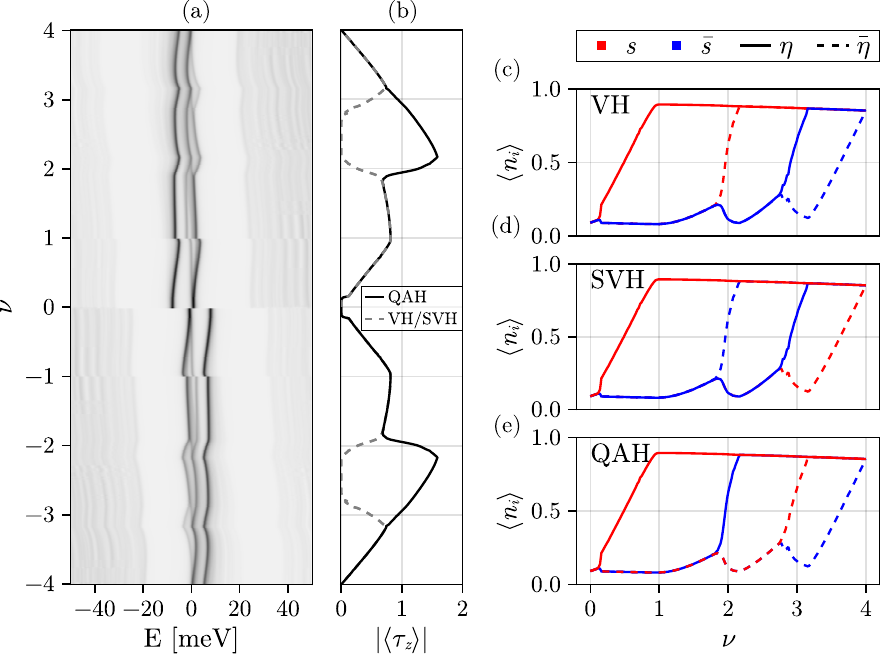}
    \caption{Sequence of mean-field ground states as a function of filling in the THFM: (a) Normalized density of states referred to the Fermi surface as a function of filling. (b) Valley polarization of $f$-electrons as a function of the filling. (c-e) Occupation of the spin (s) and valley ($\eta$) $f$ degrees of freedom vs filling for three distinct ground states corresponding to VH (c), SVH (d), and QAH (e) phases}. Red and blue colors and solid and dashed lines encode different spin and valley flavors, respectively. Parameters: $U= 5 \text{ meV}, \Delta_{1} = 2.5 \text{meV} $ and $ J= 1 \text{meV}$.% \noteHO{En la leyenda de arriba en el panel c: no deberia ser $s$ en lugar de $\sigma$ para denotar el spin (como en las Eqs. 4 en adelante)?} 
    \label{fig:enter-label}
\end{figure}

\begin{figure*}[t]
    \centering
    \includegraphics[width=1\linewidth]{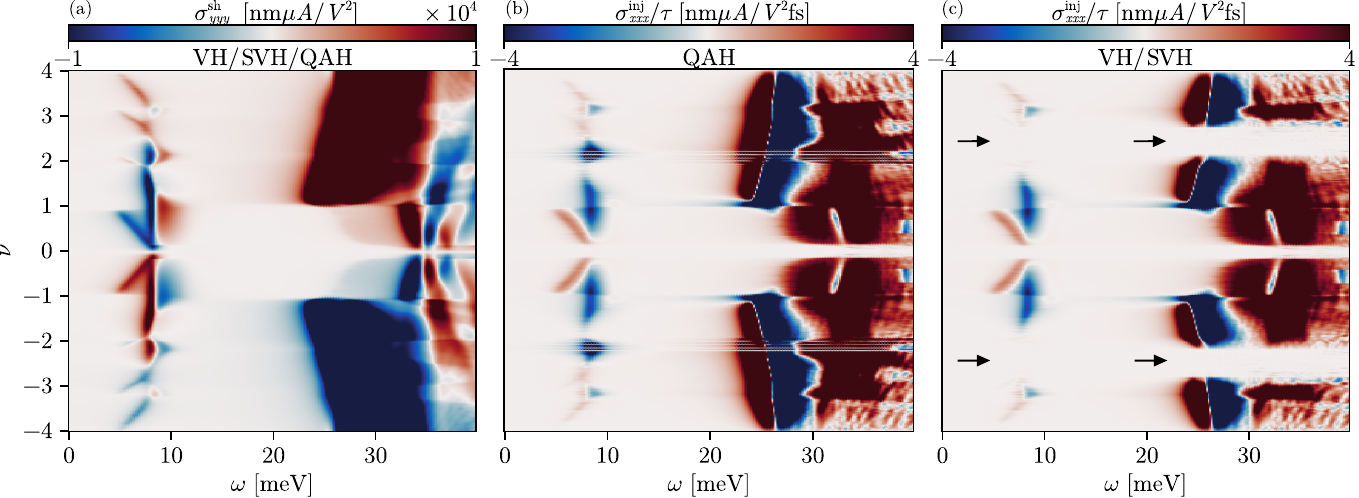}
    \caption{(a) Shift and (b,c) magnetic injection components as a function of frequency and filling in the presence of a layer-even sublattice mass $\Delta_1 = 2.5 \text{ meV}$ for the different interacting ground states described in Fig. 2. Whereas the QAH, VH, and SVH phases give rise to the same $\sigma_{yyy}^\text{shift}$, those with different Chern number: $0$ (VH/SVH) and $2$ (QAH), lead to a different $\sigma_{xxx}^\text{inj}$ response in (b) and (c) between fillings 2 and 3, respectively. Filling regions where $\sigma_{xxx}^\text{inj}$ vanishes at all frequencies are highlighted by arrows corresponding to the VH/SVH states. Same parameters as in Fig. 2.}%Parameters: $U= 5 \text{ meV}, \Delta_{\sigma_z} = 2.5 \text{meV}, J= 1 \text{meV}$.}}
    \label{fig:layerevenmass}
\end{figure*}

Remarkably, even in the presence of the sublattice effects described above, the THFM has an intrinsic spin-valley $SU(2)_K \times SU(2)_{K'}$ symmetry that makes the three polarized states: the QAH, VH, and VSH exactly degenerate.
This degeneracy within the valley polarized space is thought to be broken by Hund coupling terms which go beyond the scope of this work (see more details in Sec. \ref{conclusion}).

The $f$ electrons occupy a central role in the hierarchy of order parameters, as evidenced by the fact that the leading interacting terms in the mean-field in Eqs. \eqref{HU} and \eqref{HJ} only involve density matrices $O^f$ corresponding to the local orbitals. Therefore, the spin-valley $SU(2)_K \times SU(2)_{K'}$ symmetry can be enforced in the self-consistency by assuming diagonal density matrices of the $f$-electrons.

The results of the self-consistency are summarized in Fig.~\ref{fig:enter-label}. The existence of the periodic resets in chemical potential can be seen in the DOS as a function of filling in Fig.~\ref{fig:enter-label}(a) and the evolution of the valley polarization $|\langle \tau_z \rangle|$  for each of the three polarized phases is shown in Fig.~\ref{fig:enter-label}(b).
In order to select the VH, SVH, and QAH character within the energetically degenerate phase space, the Hartree-Fock solution is constrained by considering specific filling sequences of the diagonal density matrix of the $f$-electrons shown in Fig.~\ref{fig:enter-label}(c). In this figure, the partial occupation of spin and valley degrees of freedom is displayed as a function of filling with panel labels VH, SVH, and QAH strictly referred to $\nu = 2$.
 
Small energy deviations and chemical potential shifts of a set of narrow bands are accounted for by following adaptive $k$-integration combined with random seed initialization of the $f$-density matrix entries. Note that only the contribution of the $f$-degrees of freedom is shown, which explains why when full polarization in the valley or spin degrees of freedom is achieved, it is not accompanied by integer expectation values.

Photocurrents are computed at zero temperature from the self-consistent solutions using the length-gauge expressions in Eqs.~\eqref{shift} and \eqref{maginj}.
As follows from the symmetry analysis in Sec.~\ref{Photovoltaiceffects}, polarized states that break time-reversal symmetry should have an extra photocurrent coming from magnetic injection in Eq.~\eqref{maginj}. This is true for all odd-filling states, which realize a spin and valley polarized Chern insulator with $|C|=1$, while for half-filling it depends on the state: The QAH state does break time-reversal symmetry, while the SVH preserves it, and the VH preserves a spinless time-reversal symmetry. Magnetic injection is thus allowed in QAH and forbidden in VH and SVH phases.
Finally, for the competitive coherent states of IKS and KIVC in the absence of a sublattice mass, the magnetic injection is forbidden due to an intrinsic and an effective time-reversal symmetry, respectively.

\section{Photocurrent spectroscopy}\label{photospect}
Regarding the photocurrent responses in each of these ground states, we first consider the simplified case where the substrate induces a layer-even sublattice potential ($\Delta_1$) and the $C_{2y}$ axis is preserved. In the presence of valley polarization (which breaks $C_{2y}$ and $\mathcal{T}$ but preserves their product $C_{2y}\mathcal{T}$), magnetic injection is allowed only in $\sigma_{xxx}^{\rm inj}$ and shift in $\sigma_{yyy}^{\rm sh}$ while $\sigma_{xxx}^{\rm sh}=\sigma_{yyy}^{\rm inj}=0$ due to $C_{2y}\mathcal{T}$. This limiting case thus has the interesting feature that injection and shift currents can be distinguished by their direction. Fig. \ref{fig:layerevenmass} shows $\sigma_{xxx}^{\rm inj}$ and $\sigma_{yyy}^{\rm sh}$ as a function of filling and frequency, in a range that includes optical transitions between flat bands (FF) (5-15 meV) and between flat and dispersive bands (FD) (20-40 meV), schematically shown in Fig. \ref{fig:sketch}(c). Both photocurrents show much stronger FD transitions compared to FF transitions, as found previously without valley polarization~\cite{Kaplan22,Chaudhary22,Penaranda24}, but their detailed dependence is however different, showing several sign changes in both frequency and filling. $\sigma_{yyy}^{\rm sh}$ is insensitive to the particular polarization (QAH/VH/SVH), while $\sigma_{xxx}^{\rm inj}$ does show a dependence, vanishing between fillings 2 and 3 in the case of the VH/SVH states as expected (see the arrows in panel (c)). Fig.~\ref{fig:layerevenmass} also shows that the important constraints for the photocurrent~\cite{Penaranda24} imposed by the approximate particle-hole symmetry of TBG~\cite{Moon13,Morell17,Ahn21,Song19,Song21,Bultinck20,Bernevig21}, are realized even in the correlated states. This can be seen in panel (a), where the shift current is odd under filling reversal, and in panels (b) and (c), which show that the injection current, assuming the same sign of valley polarization through the filling cascade, is even under filling reversal.

\begin{figure*}[t]
    \hspace{-2cm}
    \includegraphics[width=1.09\linewidth,left]{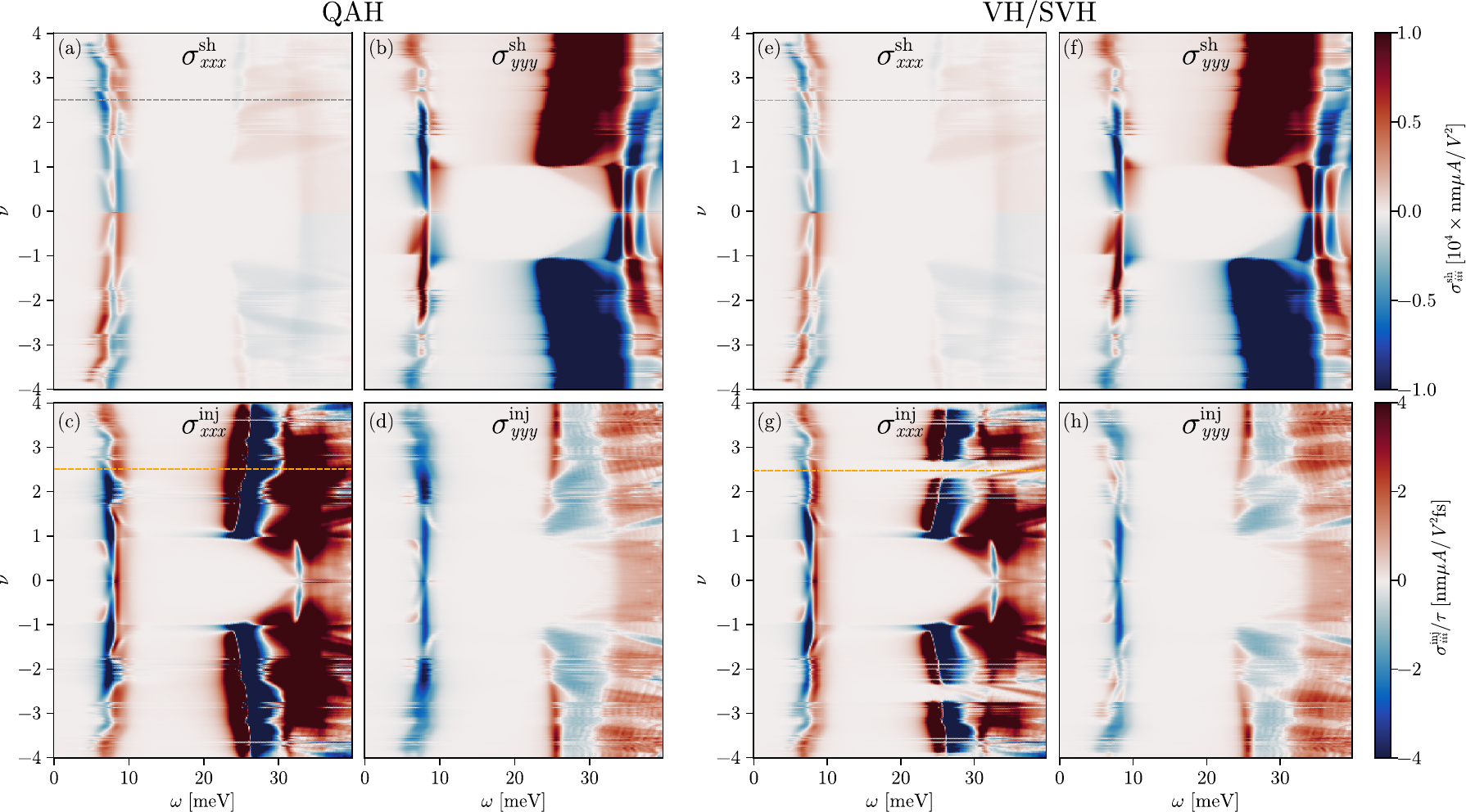}
    \caption{In-plane photocurrents in the presence of a layer-odd sublattice mass for the QAH (a-d) and VH/SVH (e-h) phases as a function of frequency and filling. (a,e) $\sigma_{xxx}^{\rm sh}$ (b,f) $\sigma_{yyy}^{\rm sh}$ (c,g) $\sigma_{xxx}^{\rm inj}$ and (d,h) $\sigma_{yyy}^{\rm inj}$. Parameters: $\Delta_{1} = 2.5 \text{ meV, } \Delta_{2} =  2.4 \text{ meV}$, the rest as in Fig.~\ref{fig:layerevenmass}.}
    \label{fig:realisticmass} 
\end{figure*}

For a more realistic account of the effect of the substrate we now study the effect of a finite layer-odd potential corresponding to the second term in Eq. \eqref{subs}. Fig. \ref{fig:realisticmass} shows both the shift and injection responses for the different polarization sequence with $\Delta_1\neq 0$ and $\Delta_2\neq0$. Since only $C_{3z}$ remains, both $\sigma_{xxx}^{\rm sh}$ (a,e) and $\sigma_{yyy}^{\rm inj}$ (d,h) become allowed. As a result and in contrast to Fig. \ref{fig:layerevenmass}, now the layer-odd mass makes the injection and shift currents not separable by directionality, having the total photocurrent contributions from both responses. However, we observe that $\sigma_{xxx}^{\rm inj}$ (c,g) and $\sigma_{yyy}^{\rm sh}$ (b,f)  remain the dominant components with similar features as in Fig. \ref{fig:layerevenmass}, while $\sigma_{yyy}^{\rm inj}$ (d,h) and $\sigma_{xxx}^{\rm sh}$ (a,e) are smaller. 

\begin{figure}[t!]
    \centering
    \includegraphics[width=1\linewidth]{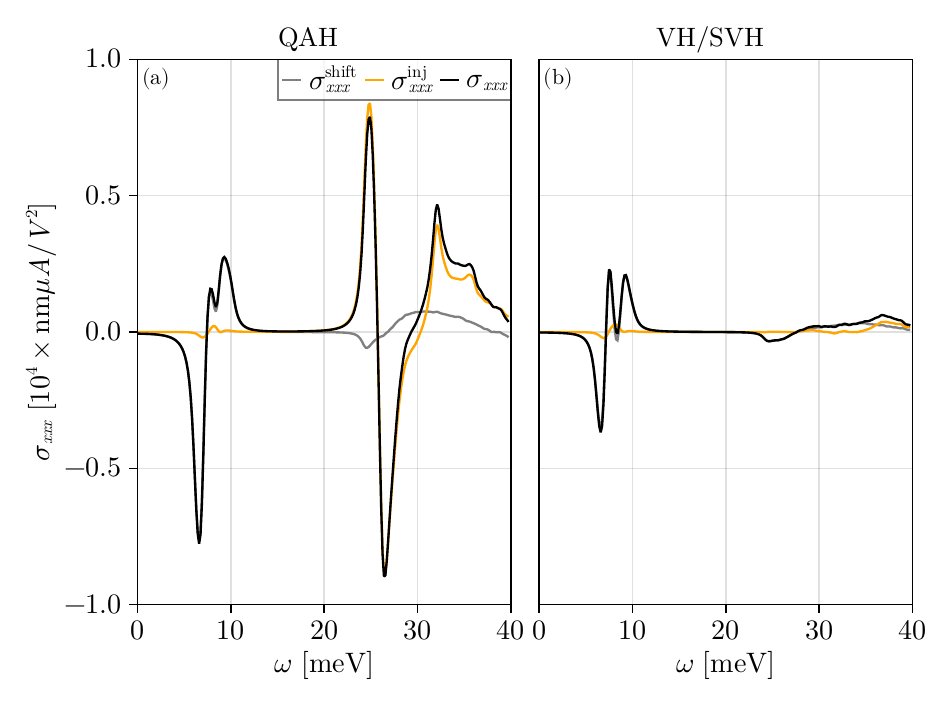}
    \caption{Shift (gray) and injection (orange) current contributions to the total photogalvanic tensor (black) along the $x$ direction. Panel (a) shows linecuts from Figs.~\ref{fig:realisticmass}(a,c), and panel (b) from Figs.~\ref{fig:realisticmass}(e,g), corresponding to the QAH and VH/SVH phases, respectively. $\tau = 150$ fs.}
    \label{fig:realisticmasslinecut}
\end{figure}

In order to make this quantitative comparison more explicit, we show in Fig. \ref{fig:realisticmasslinecut} the total current $\sigma_{xxx} =  \sigma^\text{shift}_{xxx} + \sigma^\text{inj}_{xxx}$ (black), together with the shift (gray) and injection (orange) contributions separately,  along the linecuts in Fig. \ref{fig:realisticmass}(a,c,e,g) corresponding to $\nu = 2.5$. Choosing a value of $\tau = 150$ fs, compatible with experimental measurements\cite{Monteverde10}, we observe that the injection current dominates the response due to FD optical transitions $\omega>15$ meV in the QAH phase (Fig.~\ref{fig:realisticmasslinecut}a). In addition, this valley-polarized response is one order of magnitude larger than the total current found in the VH/SVH phase (see Fig.~\ref{fig:realisticmasslinecut}b), which is purely of shift origin.

Therefore, in the more general case of Figs.~\ref{fig:realisticmass}, \ref{fig:realisticmasslinecut} with the two substrate potentials $\Delta_1$ and $\Delta_2$, we find that a direct DC-current measurement along the $x$ direction is still a good indicator of spontaneous valley polarization, not even requiring the isolation of the injection contribution by an external magnetization-reversal procedure (see the discussion in the Conclusion section).

\section{Conclusion}\label{conclusion} In our work, we have predicted a magnetic injection photocurrent in TBG which becomes enabled due to spontaneous valley polarization. The most distinctive feature of such photocurrent is that it is magnetically switchable~\cite{Zhang19}: if the magnetization can be reversed by external means, the magnetic photocurrent reverses and can be isolated as the photocurrent difference between states with opposite magnetizations. Since valley polarization is a time-odd $B_1$ irrep of $D_6$, it should be switchable with an in-plane magnetic field~\cite{Kwan20,Sharpe21,Antebi22}, which to cubic order contains the time-odd $B_1$ irrep $B_x^3-3B_xB_y^2$. Alternatively, current pulses \cite{Su20,He20,Ying21} were also shown to flip the QAH effect experimentally \cite{Sharpe19,Serlin20}. In the presence of a substrate which breaks all twofold axes, circular light can also be used to switch the QAH\cite{Persky25}, which enables an all-optical way to probe the switchability of the magnetic injection current. All of these methods can also be used to enforce the same sign of magnetization while filling is sweeped, allowing to check the prediction that the injection current is odd under filling reversal. Another key feature of the injection current is that it grows with the scattering time $\tau$ while the shift current does not, so cleaner samples should show stronger injection currents.

The injection photocurrent neatly discriminates between topological QAH and trivial VH and SVH states, which are degenerate in energy within the THFM. Their degeneracy has been proposed to be lifted by a valley-spin Hund coupling that breaks the independent spin rotational symmetry of each valley, resulting from Coulomb scattering between valleys or electron-phonon coupling with K-valley phonons~\cite{Chatterjee20}, but also by an spin-orbit-driven ferromagnetism induced by the substrate~\cite{Jiang22}. In both cases, the values of these couplings remain uncertain. In this regard, the injection current can also be used to discern the dominance of these mechanisms.

To compare our theory with the experimental observation of bulk photocurrents~\cite{Kumar25NM}, a number of caveats should be taken into account. Photocurrents have been measured both in insulating and metallic samples at neutrality~\cite{Kumar25NM}, suggesting $C_{2z}$ breaking perturbations are inhomogeneous~\cite{Shi21,Grover22} and not always lead to a global gap. This implies that photocurrent patterns will vary accross the sample, making an estimation of the sublattice potential challenging. In addition, the crystallographic direction in these experiments is not currently known, and strain effects may also play a role in the breaking of symmetries. While a more systematic study is required to isolate the injection current in the experiment, we believe the sign changes which have been observed as a function of filling are strongly indicative of a dependence on the spin-valley polarization. We hope the characterization of the magnetic injection provided here, including its frequency, filling, and polarization dependence, as well as its approximate particle-hole constraints, will serve to unambiguously establish its existence in further experiments.

\emph{Data Availability} -
Computer codes, raw data and analysis scripts for all presented figures are available in the Zenodo database under accession code: https://doi.org/10.5281/zenodo.17714403

\emph{Acknowledgements} - %We thank ... for insightful discussions. 
This work is supported by Grant PID2021-128760NB0-I00 from the Spanish MCIN/AEI/10.13039/501100011033/FEDER, EU. F.P acknowledges support from a Juan de la Cierva Fellowship (Grant No. JDC2023-051274-I) funded by MICIU/AEI/10.13039/501100011033 and the ESF+.

\bibliography{twisted}

\end{document}